\documentclass[runningheads]{llncs}
\usepackage{graphicx}
\usepackage[misc]{ifsym}
\usepackage{import}
\usepackage[utf8]{inputenc}
\usepackage{amssymb}
\usepackage{cite}
\usepackage{siunitx}
\usepackage[caption=false]{subfig}
\usepackage{lipsum}

\usepackage{multirow}
\usepackage{enumitem}
\usepackage{booktabs}
\usepackage{siunitx}
\usepackage{hyperref}
\hypersetup{
    colorlinks=true,
    linkcolor=blue,
    filecolor=magenta,      
    urlcolor=blue,
    citecolor = blue,
}

\begin{document}
\title{Decoupling Inherent Risk and Early Cancer Signs in Image-based Breast Cancer Risk Models}
\titlerunning{Decoupling breast cancer risk models}

\author{Yue Liu\inst{1,2}$^{({\textrm{\Letter}})}$ 
\and Hossein Azizpour\inst{1}  
\and Fredrik Strand\inst{3, 4}  
\and Kevin Smith\inst{1,2}} 

\authorrunning{Y. Liu et al.}

\institute{KTH Royal Institute of Technology, Stockholm, Sweden \\ \email{yue3@kth.se} \and
Science for Life Laboratory, Solna, Sweden \and
Karolinska Institutet, Stockholm, Sweden \and 
Karolinska University Hospital, Stockholm, Sweden}

\maketitle              

\begin{abstract}
The ability to accurately estimate risk of developing breast cancer would be invaluable for clinical decision-making. 
One promising new approach is to integrate image-based risk models based on deep neural networks. 
However, one must take care when using such models, as selection of training data influences the patterns the network will learn to identify.
With this in mind, we trained networks using three different criteria to select the positive training data (\textit{i.e.}~images from patients that will develop cancer):
an \textit{inherent risk} model trained on images with no visible signs of cancer, 
a \textit{cancer signs} model trained on images containing cancer or early signs of cancer, and a \textit{conflated} model trained on all images from patients with a cancer diagnosis.
We find
that these three models learn distinctive features that focus on different patterns, which translates to contrasts in performance. 
Short-term risk is best estimated by the cancer signs model, whilst long-term risk is best estimated by the inherent risk model. 
Carelessly training with all images conflates inherent risk with early cancer signs, and yields sub-optimal estimates in both regimes.
As a consequence, conflated models may lead physicians to recommend preventative action when early cancer signs are already visible.
\end{abstract}

\section{Introduction}
Breast cancer is the most commonly occurring type of cancer worldwide for women \cite{bray2018globocan}. An effective method to reduce breast cancer mortality is to detect it early while it is still curable. 
Population-wide mammographic screening is proven to have a positive effect in this regard, and has been implemented across many developed countries \cite{duffy2002impact}.
However, studies have shown that mammographic screening has limited sensitivity for some women \cite{kolb2002comparison}. 
Cancers that could potentially be found with more sensitive screening methods are routinely missed. 
For example, adding MRI or ultrasound screening would improve early detection, but are too costly to offer to the whole population.
A reliable method to estimate breast cancer risk would allow hospitals to offer more personalized care to high-risk women, including enhanced screening and other preventive measures.

\begin{figure}[t!]
\centering
\includegraphics[width=0.9\textwidth]{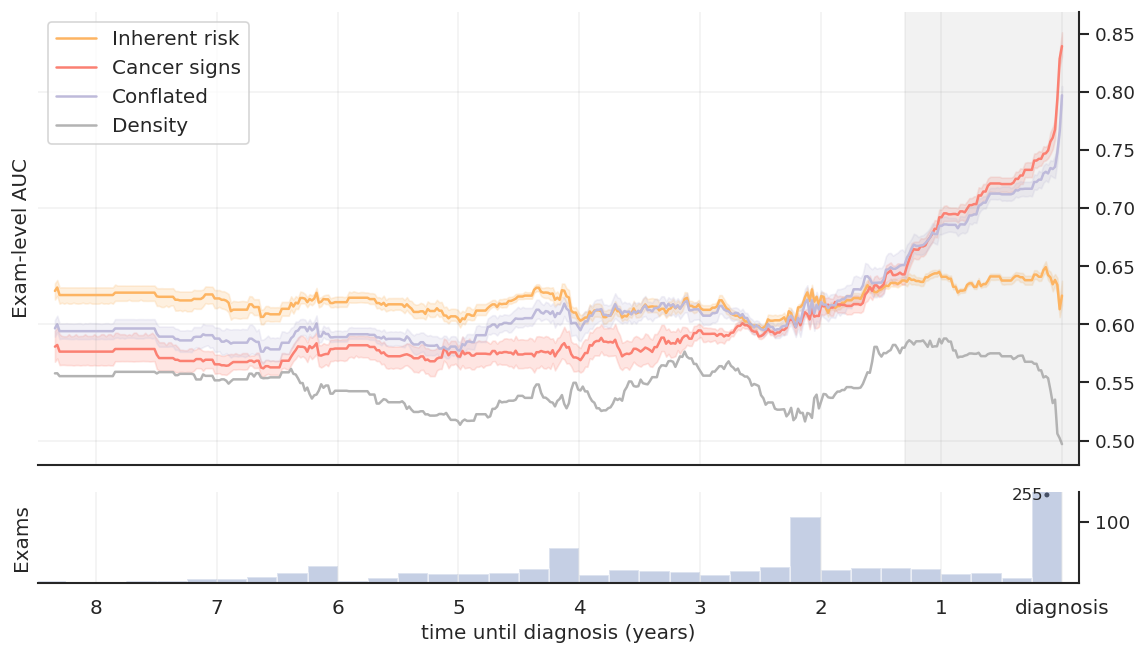}
\caption{If trained carelessly,  \textit{conflated} (blue) neural networks confound long-term \textit{inherent risk} (orange) and early \textit{cancer signs} (red). 
Top: Test AUC of the models (based on 20\% of the positive samples and all negative samples), computed every 7 days over an 8-year period using a sliding window of varying width (shaded areas show 95\% confidence interval).
Bottom: Number of positive exams in the test set (women who will develop cancer) vs.~time until diagnosis. 
Spikes correspond to scheduled screenings. 
The right-most bin contains 255 exams.  
The gray region shows when the sliding window contains samples $\leq$30 days to diagnosis, which likely corresponds to screen-detected cancers.
The \textit{conflated} model, trained on all images,
is decoupled into \textit{inherent risk} and \textit{cancer signs}.
All three models outperform the mammographic density baseline.
Short-term risk is best estimated by the cancer signs model, which is unsurprising as it was trained like a cancer detector. Long-term risk is best estimated by the inherent risk model, whose AUC remains constant even near diagnosis.
Although the conflated model was trained with more data, 
it is sub-optimal in both the short- and long-term.
} 
\label{fig:main_plot}
\end{figure}

Breast cancer risk prediction approaches include  questionnaire-based models such as Gail and Tyrer-Cuzick models \cite{gail2011personalized, tyrer2004breast} and breast density models.
A new state-of-the-art in breast cancer risk estimation was recently established using deep neural networks trained on mammograms
\cite{yala2019deep, dembrower2019comparison}. 
These risk models represent a paradigm shift towards learned features,
and have been shown to substantially outperform prior models. 
Based on these successes, we anticipate that risk assessment research will shift towards deep learning approaches.

The key message of this work is a warning that, if care is not taken when selecting the training data and designing the training procedure, \textit{neural networks trained to estimate breast cancer risk may conflate actual risk prediction and cancer detection}. 
Conflated models purport to perform long-term risk prediction, but in reality are highly sensitive to cancer signs. 
This yields sub-optimal long-term risk estimation, and could cause cancers to go undetected if physicians believe women have high long-term risk when in fact they exhibit cancer signs. 

Through a series of experiments, we illustrate the phenomenon of risk conflation both qualitatively and quantitatively, and measure how it impacts the performance of risk prediction over time.
Code to reproduce our work is available at \url{https://github.com/yueliukth/decoupling_breast_cancer_risk}. 

\section{Related Works}
Breast cancer prevention demands accurate and individualized risk assessment for decision-making. Over the last decades, many models for estimating individual breast cancer risk have been developed.
The Gail model \cite{gail2011personalized} is a questionnaire-based method for estimating 5-year and lifetime risk of developing invasive breast cancer.
It considers risk factors such as a woman's age and family history.
Tyrer–Cuzick \cite{tyrer2004breast}, another commonly-used risk model, incorporates more detailed family history.
Glynn \textit{et al.} recently compared questionnaire-based models and found that their practical usefulness is limited by performance \cite{glynn2019comparison}.

Breast density, aside from age, is one of the strongest risk factors for breast cancer \cite{boyd2007mammographic}. 
Density measures if a breast is more fatty or contains more fibroglandular tissue, can be obtained from mammographic screens, and has been shown to improve questionnaire-based models \cite{brentnall2015mammographic}.
Density is often defined by a few statistics obtained either through ad-hoc \cite{rauh2012percent} or learning-based approaches \cite{keller2012estimation}. 
In general, methods for quantifying density lack consistency \cite{amir2010assessing} and tend to over-simplify image data, limiting their general application.

In the era of deep learning, most research in mammography has focused on computer-aided diagnosis (CAD) \cite{geras2017high, shen2019deep, mckinney2020international}. 
A handful of studies have addressed risk prediction, though most have been restricted to small datasets and short-term prediction.
Two such studies \cite{sun2016preliminary, qiu2016initial} considered a few hundred negative screening samples, 
and predicted which would be positive at the next screening.
He \textit{et al.} used a multi-modal approach to combine mammographic screenings, ultrasound images, patient demographics, and language from clinical reports to predict if a patient with an abnormal mammogram should be sent for biopsy \cite{he2019deep}.

Two recent breakthrough studies showed substantial improvements in long-term risk prediction using neural networks on large population-level cohorts.
Yala \textit{et al.} showed mammogram-based deep learning models 
outperform the Tyrer–Cuzick model for five-year risk prediction \cite{yala2019deep}.
Dembrower \textit{et al.} similarly showed that five-year risk predictions from a neural network surpass density-based predictions \cite{dembrower2019comparison}. 
In this study, we consider the same cohort as Dembrower \textit{et al.}, but our focus is not to push performance, rather to raise awareness of the dangers of conflating long-term risk and cancer signs in risk models.

\begin{figure}[t]
\centering
\begin{tabular}{@{}l@{}c@{\hspace{2mm}}c@{}}
 & \hspace{-5mm} \small{Positive samples for \textit{inherent risk}} & \hspace{-5mm} \small{Positive samples for \textit{cancer signs}} \\
\rotatebox{90}{\hspace{2mm} \scriptsize{\textsf{ipsilateral}}} &
\includegraphics[width=0.48\textwidth]{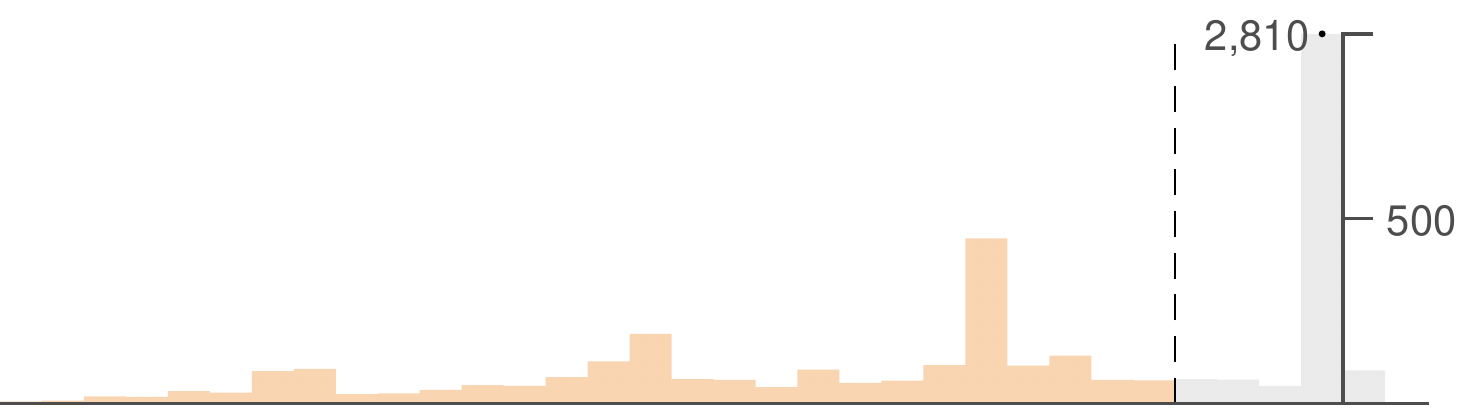} &
\includegraphics[width=0.48\textwidth]{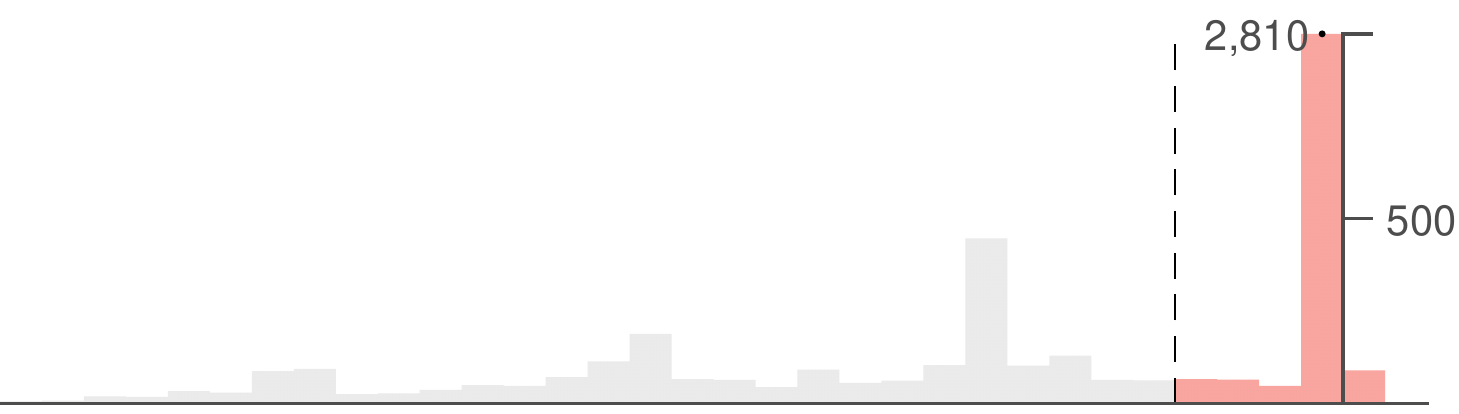} \\
\rotatebox{90}{\hspace{6mm} \scriptsize{\textsf{contralateral}}} &
\includegraphics[width=0.48\textwidth]{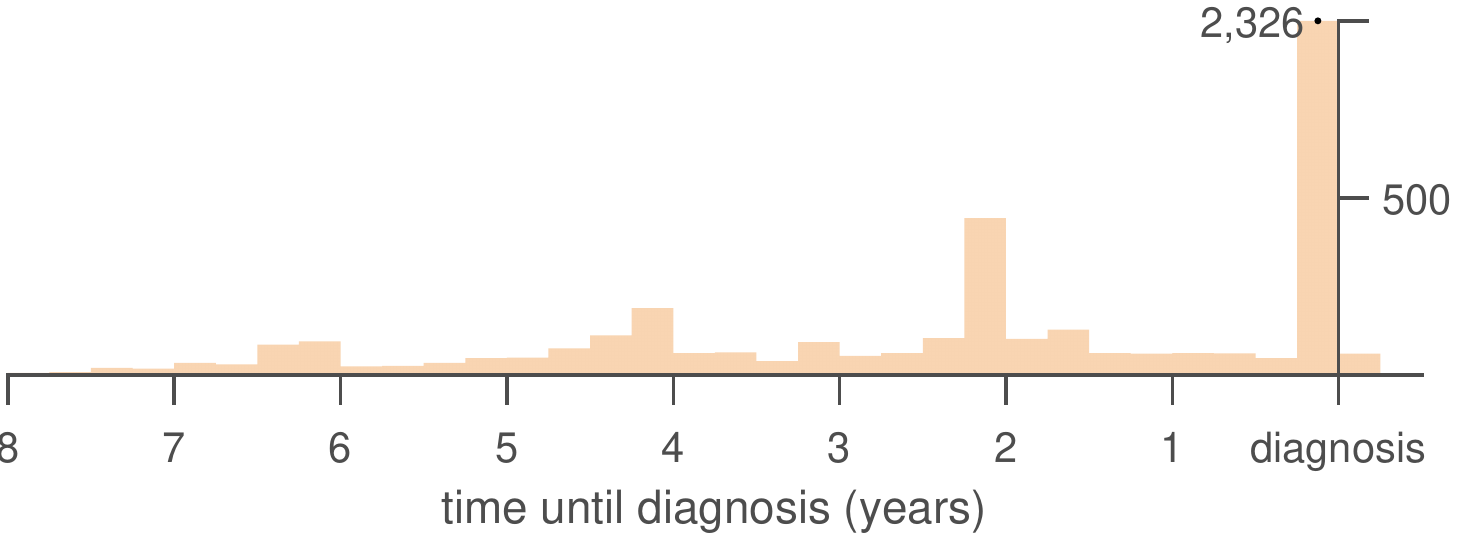} &
\includegraphics[width=0.48\textwidth]{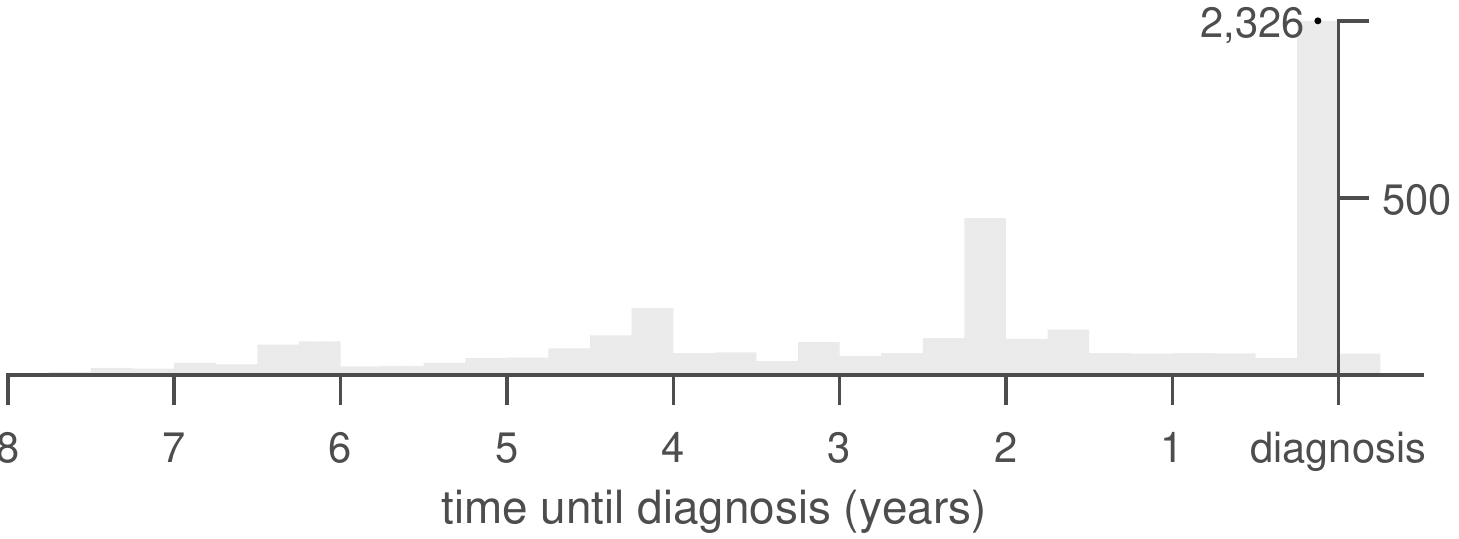} \\
\end{tabular}
\caption{We decouple \textit{inherent risk} (orange) and \textit{cancer signs} (red) from the \textit{conflated} model by splitting the positive training data in two separate parts. Top: Histogram of positive ipsilateral images over the study period (ipsilateral is the the breast that develops cancer). A cutoff of 1 year from diagnosis (dashed line) separates images with no visible cancer signs (orange) from those with possible cancer signs (red). Bottom: Positive contralateral images, from the other breast that is usually confirmed cancer-free. As it has been exposed to the same environmental and genetic risk factors as the ipsilateral, it is included in the inherent risk model. The conflated risk model is trained with all images. The cancer signs model is trained with red-marked positive examples, while the inherent risk model is trained with orange-marked positive examples. } 
\label{fig:dataset}
\end{figure} 

\section{Decoupling Breast Cancer Risk}
A straightforward approach to train a network to predict breast cancer risk from mammograms is to provide all images from cancer patients as positive examples.
Several prior works have trained models in this manner. 
The problem with this approach is that the images recorded near the date of diagnosis are included in the positive set, and are likely to include signs of actual cancer.
We can imagine separating the positive training images with no visible cancer signs from those containing cancer signs by drawing an arbitrary cutoff within one screening interval, \textit{e.g.} at one year from diagnosis (Fig. \ref{fig:dataset}). 
From this perspective, the data contains two different classification problems: \textit{inherent risk vs.~healthy} and \textit{cancer signs vs.~healthy}. When we train using all the data, we conflate them into a single binary classification task, \textit{at-risk vs.~healthy}.

This is problematic for the learning process, since recognizing long-term risk is more difficult than detecting cancer signs. 
Networks are known to converge faster with easier examples~\cite{bengio2009curriculum,weinshall2018curriculum}, and if it focuses too strongly on increasing confidence of the easy samples~\cite{guo2017calibration} learning on the harder long-term risk problem may be crippled.

Consequently, we hypothesize that the conflated model will perform worse at long-term risk prediction than a model trained exclusively with images acquired before onset of early cancer signs.
This effect will be more acute when a substantial portion of the positive data contains cancer signs, which is typical for population datasets (for CSAW \cite{dembrower2019multi}, up to 31\% of the positive samples may contain cancer signs).
In order to test this hypothesis, we decomposed the conflated model by dissecting the data and training models on those splits. We trained an \textit{inherent risk} model using data with no visible cancer signs, and a \textit{cancer signs} model using data that contains a substantial number of cancer signs. The \textit{conflated} model was provided with all available data.

Details of the data selection strategy are provided in Fig.~\ref{fig:dataset}. 
For ipsilateral -- breasts that will develop cancer --
we selected a cutoff of one year prior to diagnosis to separate inherent risk and cancer signs (dashed line). 
The contralateral breast is usually confirmed cancer-free in patients with breast cancer. 
It reflects actual risk without revealing any cancer cues, as it has been exposed to the same environmental and genetic risk factors. 
Therefore, we included the contralateral breast in the inherent risk model but not the cancer signs model.

Using these models, we conducted a series of experiments to understand the phenomenon of risk conflation. 
We address the following questions:
\begin{enumerate}
\item \textit{How does the conflated model compare to the decoupled models over time?}
\item \textit{Does the conflated model identify the same at-risk women as the inherent risk/cancer signs models?} 
\item \textit{Do the inherent risk/cancer signs models recognize the same patterns?}
\end{enumerate}

\begin{table}[t]
\caption{Model Performance of Risk Prediction on Test Set}\label{tab1}
\begin{center}
\scriptsize
\begin{tabular}{@{}r@{\hspace{5mm}}l@{\hspace{1mm}}l@{\hspace{3mm}}l@{\hspace{1mm}}l@{\hspace{3mm}}l@{\hspace{1mm}}l@{\hspace{3mm}}l@{\hspace{1mm}}l@{}}
\toprule
\multirow{2}{*}{} &  \multicolumn{8}{c}{AUC (95\% CI)} \\ 
& \multicolumn{2}{c}{31d -- 1 year}  & \multicolumn{2}{c}{\textgreater 1 year} & \multicolumn{2}{c}{\textgreater 2 years} & \multicolumn{2}{c}{\textgreater 5 years} \\ \midrule
Inherent risk & 0.62 &(0.62, 0.63) & \textbf{0.62}& (0.61, 0.62) & \textbf{0.62}&(0.61, 0.62) &
\textbf{0.61} & (0.60, 0.62)\\
Cancer signs & \textbf{0.71} &(0.68, 0.73)  & 0.59 & (0.58, 0.60) & 0.59 & (0.58, 0.59) & 0.56 & (0.55, 0.57)\\
Conflated & \textbf{0.72} & (0.69, 0.75) & \textbf{0.61} & (0.60, 0.62) & 0.60 & (0.59, 0.61) & 0.58 & (0.56, 0.59)\\
Density & 0.61 & (N/A) & 0.54 & (N/A)  & 0.54 & (N/A) & 0.55 & (N/A)\\
\bottomrule
\normalsize
\end{tabular}
\end{center}
\end{table}

\section{Experimental Setup}
\subsubsection{Dataset}
The dataset used in our study is extracted from CSAW, a population-based screening cohort containing millions of mammographic images \cite{dembrower2019multi}.
Mammograms of multiple views were collected every 18 to 24 months from women aged 40 to 74. 
Outcome and date of diagnosis was determined through the Regional Cancer Center Registry. 
The data was curated by excluding images from patients with implants, biopsy images, or other issues such as aborted exposure.
We randomly assigned the participants to the training, validation and test set.
Negative exams were randomly sampled among women with at least two years' cancer-free follow-up.
A flowchart describing the data curation is given in Supplementary Figure 1.
The resulting training set contains {138,032} mammograms from {15,558} women, the validation set contains {3,008} mammograms from {332} women, and the test set contains {6,436} mammograms from {731} women. 

\subsubsection{Preprocessing}
The source images are in standard DICOM format. 
Using DICOM metadata, we flip images horizontally to make all breasts left-posed.
We rescale the intensity to the range defined in the acquisition metadata \cite{clunie2003dicom}, and we detect and correct inverted contrast images using the photometric attribute. 
We perform a rough alignment of each image using a distance transform to locate the center of mass.
Zero-padding is applied to ensure all images have uniform size, then images are resized to $632\times512$. 
This ensures each image retains relative scale and aspect ratio.
Finally, the images are converted to 16-bit PNG format. 

\subsubsection{Implementation details}
We use the same architecture and training setup for all models. 
In particular, we use ResNet50 \cite{he2016deep} with group normalization \cite{wu2018group}, and replace standard ReLU activation with Leaky-ReLU \cite{maas2013rectifier}.
We use binary cross-entropy loss and batch size of 32 with a stochastic gradient descent with momentum (SGDM) optimizer.
All models were initialized with ImageNet pretrained weights \cite{deng2009imagenet}.
We employ standard data augmentation including random rotation, crops, brightness and contrast.
Hyperparameters detailed below were selected using grid search.
The initial learning rate for the cancer signs and conflated models is 0.0001, and 0.001 for the inherent risk model.
The inherent risk and conflated models were run for 50 epochs, and the learning rate was lowered by a factor of 10 at epoch 20.
The cancer signs model was run for 100 epochs, with a similar learning rate drop at epoch 50.
Dropout \cite{hinton2012improving} with a rate of 0.5 was applied after the last fully connected layer in the inherent risk model.

We repeated each experiment five times and report the mean, unless otherwise specified.
As a baseline, we provide risk estimation results using mammographic density (breast dense area) from publicly available software, LIBRA \cite{keller2015preliminary}.

\begin{figure}[t]
\centering
\subfloat[\textgreater 2 years\label{fig:top5_circle1}]{\includegraphics[width=0.28\columnwidth]{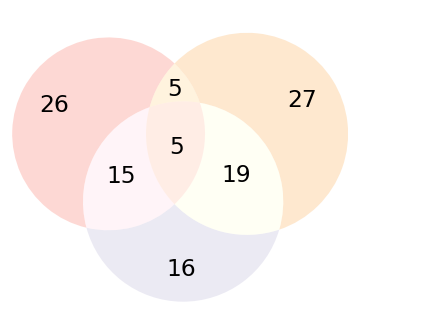}}\quad
\subfloat[31 days to 2 years\label{fig:top5_circle2}]{\includegraphics[width=0.28\columnwidth]{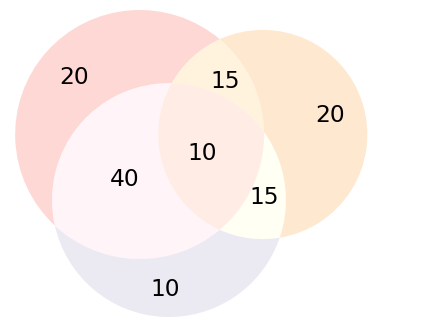}}\quad
\subfloat[\textless 31 days\label{fig:top5_circle3}]{\includegraphics[width=0.28\columnwidth]{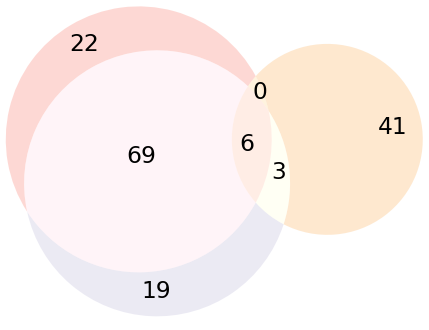}}\quad
\caption{Venn diagrams showing \textit{true positive rates}
of various models, given their top-5\% predictions.
The \textit{inherent risk} (orange) model consistently identifies decidedly different sets of at-risk women, compared to the \textit{cancer signs} (red) model. The \textit{conflated} (blue) model identifies nearly the exact same positive images as the cancer signs model near the date-of-diagnosis. But farther from diagnosis it overlaps both decoupled models.}
\label{fig:top5percent}
\end{figure}

\begin{figure}[t!]
\scriptsize
\centering
\begin{tabular}{@{}c@{\hspace{0.5mm}}c@{\hspace{.5mm}}c@{\hspace{.5mm}}c@{\hspace{1mm}}@{\hspace{1mm}}c@{\hspace{.5mm}}c@{\hspace{.5mm}}c@{}}
&
t.t.d. = 4.24  &
t.t.d. = 2.11 &
t.t.d. = 0.02 &
t.t.d. = 4.24 &
t.t.d. = 2.11 &
t.t.d. = 0.02 \\[1mm]
&
$\hat{y}$ = 0.53&
$\hat{y}$ = 0.46&
$\hat{y}$ = 0.51&
$\hat{y}$ = 0.09&
$\hat{y}$ = 0.53&
$\hat{y}$ = 0.94\\
\rotatebox{90}{\hspace{4mm} CC view} &
\includegraphics[width=0.15\columnwidth]{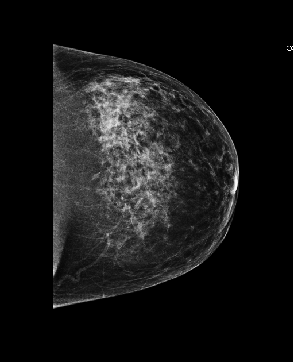}&
\includegraphics[width=0.15\columnwidth]{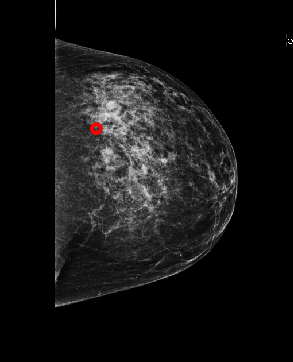}&
\includegraphics[width=0.15\columnwidth]{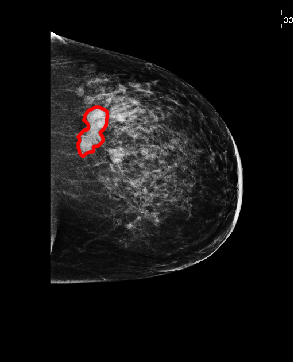}&
\includegraphics[width=0.15\columnwidth]{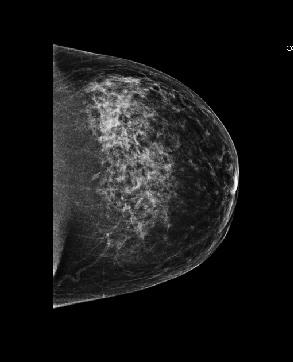}&
\includegraphics[width=0.15\columnwidth]{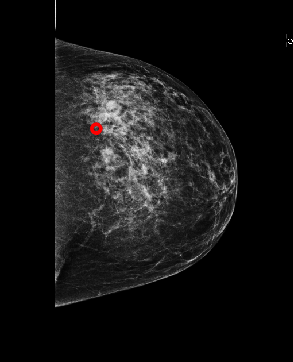}&
\includegraphics[width=0.15\columnwidth]{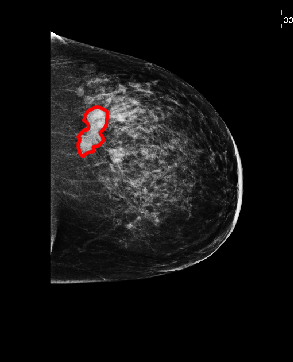}\\[-.5mm]
\rotatebox{90}{\hspace{3mm} Grad-CAM} &
\includegraphics[width=0.15\columnwidth]{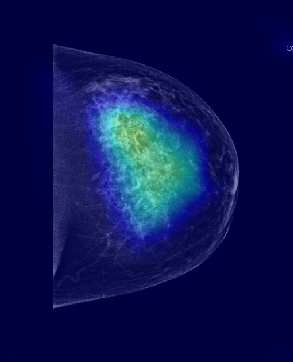}&
\includegraphics[width=0.15\columnwidth]{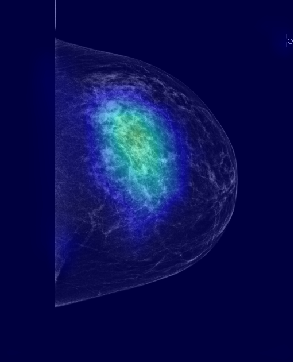}&
\includegraphics[width=0.15\columnwidth]{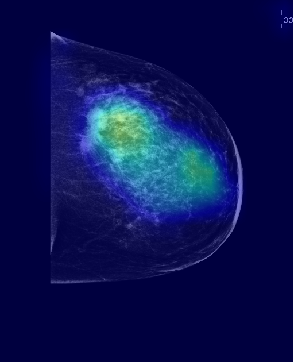}&
\includegraphics[width=0.15\columnwidth]{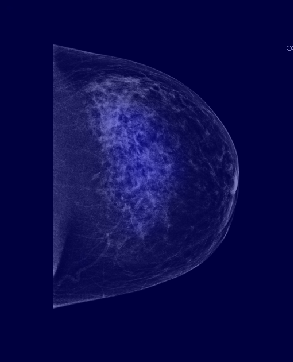}&
\includegraphics[width=0.15\columnwidth]{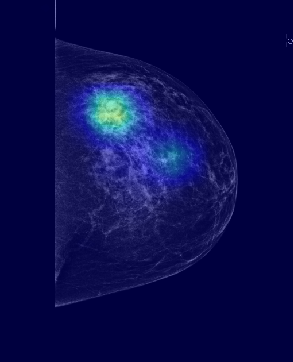}&
\includegraphics[width=0.15\columnwidth]{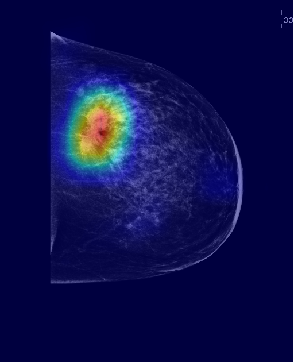}\\
& \multicolumn{3}{c}{\small (a) Inherent risk model}&\multicolumn{3}{c}{\small (b) Cancer signs model}
\end{tabular}
\caption{Grad-CAM visualizations suggest that the inherent risk model and cancer signs model base their decisions on different image cues.
Top: CC views of a breast that develops cancer over 4 years.  
An expert cancer annotation (red region) appears in the most recent image, and the cancer-developing location is identified in the prior image  (red dot).
Time-to-diagnosis (t.t.d., in years) and risk prediction $\hat{y}$ are provided for each image.
Bottom: Grad-CAM visualizations.
The activation maps are weighted by the prediction score. 
The inherent risk model appears to rely on a broad range of image cues, while the cancer signs model concentrates activations on tumor-like patterns. }
\label{fig:gradcam} 
\end{figure}

\section{Results and Discussion}
Through a series of experiments based on the setup described above, we address the questions raised in Section 3.
\subsubsection{Conflated risk model vs.~decoupled models }
We find that the conflated model is a weakened hybrid of the inherent risk and cancer signs models.
It underperforms the decoupled models in both short- and long-term risk prediction.
In Fig.~\ref{fig:main_plot} we plot the exam-level AUC for our three models along with the density baseline. 
The $x$-axis shows how performance varies with time-to-diagnosis using a sliding window.
Exam-level predictions are the maximum breast risk score; breast scores are the average score of both views.
Near diagnosis, the cancer signs model is the best risk estimator. This is unsurprising because it was trained like a tumor detector, and many of the positive mammograms within the first year, especially within the first 30 days, are screen-detected cancers with visible tumors. 
Long-term risk is best estimated by the inherent risk model, whose AUC remains constant, even in the first year.
This suggests that the inherent risk model has the desirable property of \textit{ignoring early cancer signs\footnote{Cancer detection is the purview of established screening routines or CAD systems.} and focusing on cues correlated with long-term risk, which do not change near time-of-diagnosis}.

Similar conclusions can be drawn from Table~\ref{tab1}, where we break down risk prediction by short-term and long-term outlooks. Inherent risk performs best at long-term risk prediction, while the cancer signs/conflated models show similar performance in the short-term (bold values indicate significant improvements; statistical tests can be found in Supplementary Table 1).

\begin{figure}[!ht]
\centering
\includegraphics[width=0.82\textwidth]{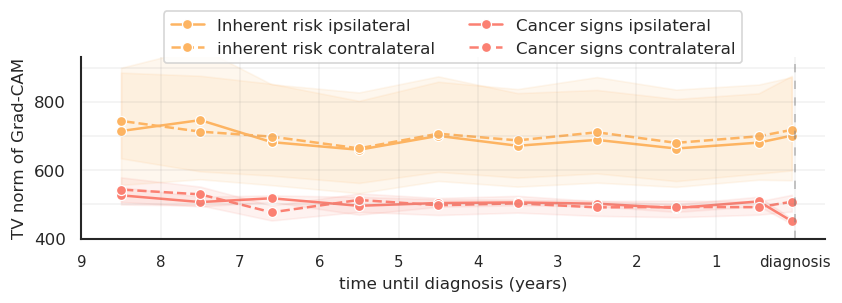}
\caption{Localization of gradient-weighted class activation maps (Grad-CAM) from Fig.~\ref{fig:gradcam}, computed over all positive test images ($x$-axis indicates time until diagnosis).
Lines are the mean computed from five models, shaded areas indicate the 95\% CI.
The cancer signs (red) model exhibits more localized activations, measured by total variation (TV) norm of Grad-CAMs.
The inherent model (orange) consistently covers a larger area, indicating that the two models concentrate on different patterns.
A sharp dip near time-of-diagnosis in the ipsilateral cancer signs model suggests that it concentrates on tumor-like patterns, as these images are likely to contain visible cancer.
} 
\label{fig:tv_norm}
\end{figure} 

\subsubsection{Identifying at-risk women}
An important clinical question is: do these models identify the same at-risk women?
To test this, we consider images identified by the top-5\% predictions of each model -- a number chosen to reflect the capacity of a healthcare system for additional screening.
In Fig.~\ref{fig:top5percent} we compare positive-identified images from all three models.
The inherent risk model consistently identifies different images than the cancer signs model, supporting our hypothesis that it focuses on different cues.
Near the date-of-diagnosis, the conflated model highly overlaps with the cancer signs model, but farther from diagnosis it overlaps both decoupled models. Its proportion of novel at-risk findings is consistently low, suggesting it could be replaced by the decoupled models.

\subsubsection{Image cues that indicate risk}
The final question we address is: do the decoupled models recognize different patterns?
This is a difficult question to answer conclusively, but we can gain some insight by
understanding and quantifying where the network pays attention.

In Fig.~\ref{fig:gradcam} we visualize how gradient-weighted class activation maps (Grad-CAM) \cite{selvaraju2017grad} of the inherent risk and cancer signs models evolve over time.
Qualitatively, we can see that the cancer signs model exhibits sharp activations localized to the tumor, whereas the inherent risk model has broad activations in the center of the breast.
We empirically confirm this trend over the entire positive test set by computing the total variation of the Grad-CAM heatmaps in Fig.~\ref{fig:tv_norm}, and using multi-scale blob detection \cite{lindeberg1998feature} in Supplementary Figure 2. 
Based on these results, we surmise that the inherent risk model relies on a broader range of image cues than the cancer signs model, which appears to concentrate activations near tumor-like patterns. 
\section{Conclusions}
Our key finding is that risk estimation models conflate inherent risk and cancer signs if care is not taken during training. 
We demonstrate that conflated models can be decoupled by selecting appropriate training data, and that the decoupled models consistently outperform the conflated model, even though it is trained with more data. 
In particular, short-term risk ($\leq$1 year in our study) should rely on cancer sign models. Long-term risk models should be trained exclusively on images with no visible cancer signs, or use other strategies to mitigate model conflation.
When models are put to clinical use, it is important to state which type of model is used, or to somehow assist in the interpretation of conflated models -- otherwise physicians may believe that a woman has high long-term risk when, in fact, her images already exhibit cancer signs. 
Our hope is that this work will provide valuable insights for the development and clinical translation of deep neural networks for cancer risk estimation.
\subsubsection{Acknowledgements}
This work was partially supported by Region Stockholm HMT 20170802, MedTechLabs (MTL), the Swedish Innovation Agency (Vinnova) 2017-01382, the Wallenberg Autonomous Systems Program (WASP), and the Swedish Research Council (VR) 2017-04609. 
\bibliographystyle{splncs04}
\bibliography{paper2196_references.bib}
\newpage
\appendix
\section{Appendix}
\begin{figure}[!ht]
\centering
\includegraphics[width=0.85\textwidth]{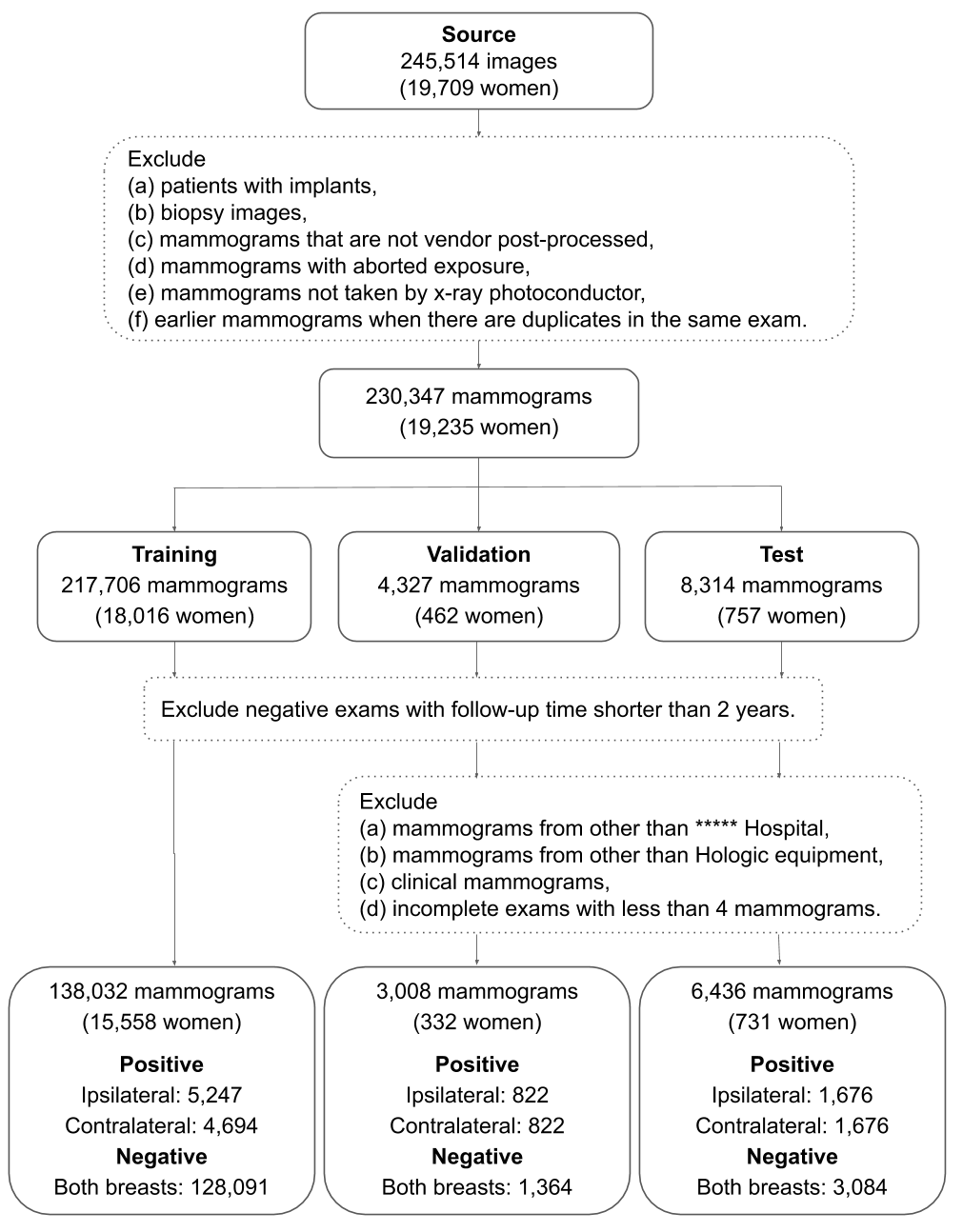}
\caption{Data selection flowchart. There were {245,514} images from {19,709} women in the source data extracted from CSAW, which included all positive cases and a random sampling of negative cases. We excluded patients with implants, biopsy images and images with other issues. After randomly splitting the remaining images into train/validation/test sets, we excluded risk-negative exams without a cancer-free follow-up screening within two years. The resulting dataset includes {138,032} mammograms from {15,558} women,  {3,008} mammograms from {332} women, {6,436} mammograms from {731} women in the training/validation/test sets.}
\label{appendix:dataset_flowchart}
\end{figure} 

\begin{table}[!ht]
\caption{Statistical tests for differences between model predictions}\label{tab2}
\begin{center}
\footnotesize
\begin{tabular}{@{}r@{\hspace{7mm}}l@{\hspace{5mm}}l@{\hspace{5mm}}l@{\hspace{5mm}}l@{}}
\toprule
\multirow{2}{*}{} &  \multicolumn{4}{c}{$p$-value  (two-sided t-test)} \\ 
& {31d -- 1 year}  & {\textgreater 1 year} & {\textgreater 2 years} & {\textgreater 5 years} \\ \midrule
Conflated / inherent risk & \textless 0.001 & 0.063 & 0.007 & 0.002\\
Conflated / cancer signs & 0.546 & 0.020 & 0.014 & 0.101\\
Cancer signs / inherent & \textless 0.001 & \textless 0.001 & \textless 0.001 & \textless0.001\\
\bottomrule
\normalsize
\end{tabular}
\end{center}
\end{table}

\begin{figure}[!ht]
\centering
\includegraphics[width=0.92\textwidth]{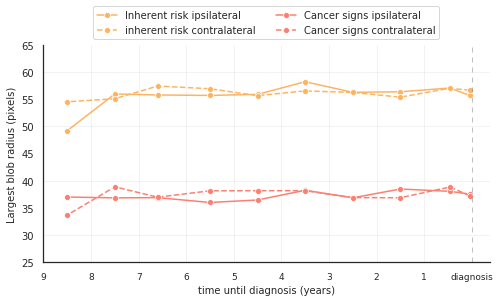}
\caption{Multi-scale blob detection on gradient-weighted class activation maps (Grad-CAM) from Fig.~4 in the main text, computed over all positive test images ($x$-axis indicates time until diagnosis).
Using highest response of a Laplacian-of-Gaussian filter at multiple scales, the radius of the largest detected blob from each positive image was recorded. 
Activations from the inherent risk model are consistently larger, suggesting it focuses on a broader range of features than the cancer signs model.} \label{fig:blob}
\end{figure} 

\clearpage
\end{document}